\documentstyle[aps,prb,multicol,floats,epsf]{revtex}
\begin{document}
\draft
\preprint{}

\wideabs{  

\title{Electronic states in the antiferromagnetic phase of electron-doped high-$T_c$ cuprates}
\author{T. Tohyama and S. Maekawa}

\address{Institute for Materials Research, Tohoku University,
        Sendai 980-8577, Japan}
\date{Received 4 September 2001}
\maketitle

\begin{abstract}

We investigate the electronic states in the antiferromagnetic (AF) phase of electron-doped cuprates by using numerically exact diagonalization technique for a $t$-$t'$-$t''$-$J$ model.  When AF correlation develops with decreasing temperature, a gaplike behavior emerges in the optical conductivity.  Simultaneously, the coherent motion of carriers due to the same sublattice hoppings is enhanced.  We propose that the phase is characterized as an AF state with small Fermi surface around the momentum {\bf k}=($\pi$,0) and (0,$\pi$).  This is a remarkable contrast to the behavior of hole-doped cuprates.
\end{abstract}

}  

\narrowtext

High-$T_c$ superconductivity emerges with carrier doping as insulating cuprates, i.e., Mott insulators.  In electron-doped cuprates such as Nd$_{2-x}$Ce$_x$CuO$_4$, an antiferromagnetic (AF) phase remains up to the concentration $x$$\sim$0.15, in contrast with hole-doped cuprates such as La$_{2-x}$Sr$_x$CuO$_4$ in which the AF phase disappears with an extremely small amount of $x$.\cite{Takagi}  On the other hand, the superconducting order parameter of the electron-doped cuprates is the same as that of the hole-doped ones: Phase-sensitive experiments,\cite{Tsuei} microwave measurements\cite{Kokales,Prozorov}, and angle-resolved photoemission spectroscopy\cite{Armitage,Sato} (ARPES) have reported $d$-wave symmetry.  This suggests the same mechanism of superconductivity in both types of cuprates.  In order to clarify common factors between the two cuprates, it is crucial to understand the nature of AF phases adjacent to superconducting phases and also to find key factors characterizing the AF phases.

Recently, the charge dynamics for AF samples of electron-doped Nd$_{2-x}$Ce$_x$CuO$_4$ has been measured and a gaplike feature has been observed in the optical conductivity $\sigma(\omega)$.\cite{Onose1,Onose2}  The gap energy increases with decreasing $x$ from 0.2~eV for $x$=0.125 to 0.45~eV for $x$=0.05.  Similar doping dependence is seen in the gap-opening temperature $T^*$ whose energy scale is one order of magnitude smaller than that of the gap: $T^*$$\sim$200~K for $x$=0.125 and $\sim$450~K for $x$=0.05.  No similar gap features have been observed in hole-doped cuprates.  Therefore a remarkable difference of the electronic states between the electron- and hole-doped cuprates is expected.

In this report, we theoretically examine the electronic and magnetic properties of electron-doped cuprates and clarify the nature of the AF phase through comparison with hole doping.  In contrast to other theoretical works\cite{FLEX} that are based on approaches from overdoped regions emphasizing spin fluctuations as well as Fermi surface nesting, we approach this problem from the Mott insulator side using a $t$-$J$ model with long-range hoppings ($t'$ and $t''$).  From previous studies,\cite{Tohyama1,Gooding} it is known that AF correlation strongly survives in the electron-doped $t$-$t'$-$t''$-$J$ model upon doping at zero temperature.  We find from finite-temperature calculations for small clusters that, when the AF correlation develops with decreasing temperature, a gaplike behavior emerges in $\sigma(\omega)$.  There is no such gap structure in the hole-doped $t$-$t'$-$t''$-$J$ model where AF correlation is strongly suppressed upon doping.  In addition to the gap feature, a large Drude weight is obtained in the small cluster, indicating coherent motion of the carriers.  It is also known that a doped electron at half filling enters into the {\bf k}=($\pi$,0) or (0,$\pi$) points in the $t$-$t'$-$t''$-$J$ model\cite{Tsutsui} in contrast to a doped hole that goes into ($\pi$/2,$\pi$/2).  The spectral function calculations indicate that doped electrons continue to occupy around ($\pi$,0) as long as the AF order persists.  Therefore we propose that the AF phase in the electron-doped cuprates is characterized as an AF state with small Fermi surface around ($\pi$,0).

The $t$-$J$ Hamiltonian with long-range hoppings, termed
the $t$-$t'$-$t''$-$J$ model, is
\begin{eqnarray}
H&=& J\sum\limits_{\left<i,j\right>_{1{\rm st}}}
      {{\bf S}_i}\cdot {\bf S}_j
    -t\sum\limits_{\left<i,j\right>_{1{\rm st}} \sigma }
    c_{i\sigma }^\dagger c_{j\sigma } \nonumber \\
&& {} -t'\sum\limits_{\left<i,j\right>_{2{\rm nd}} \sigma }
    c_{i\sigma }^\dagger c_{j\sigma }
     -t''\sum\limits_{\left<i,j\right>_{3{\rm rd}} \sigma }
    c_{i\sigma }^\dagger c_{j\sigma }+{\rm H.c.}\;,
\label{H}
\end{eqnarray}
where the summations $\left< i,j \right>_{1{\rm st}}$,
$\left< i,j \right>_{2{\rm nd}}$, and $\left< i,j \right>_{3{\rm rd}}$
run over first, second, and third nearest-neighbor (NN) pairs, respectively.  No double occupancy is allowed, and the rest of the notation is standard.  In the model, the difference between hole and electron carriers is expressed as the sign difference of the hopping parameters:\cite{Tohyama1} $t$$>$0, $t'$$<$0, $t''$$>$0 for hole doping and $t$$<$0, $t'$$>$0, $t''$$<$0 for electron doping.  The ratios $t'$/$t$ and $t''$/$t$ are estimatied to be $-$0.34 and 0.23, respectively,\cite{Kim} although they are material dependent.\cite{t'}  Here $J$/$|t|$ is taken to be 0.4.  In order to examine the model, we use numerically exact diagonalization methods for small-size clusters with $N$=$\sqrt{20}$$\times$$\sqrt{20}$ and 4$\times$4 sites.  A standard Lanczos techunique is employed at $T$=0, and for $T$$>$0 a finite-temperature Lanzos method\cite{Jaklic} is used.

Let us start with a single carrier in the Mott insulators.  The energy dispersion of the carrier is approximately given by\cite{Tohyama2}
\begin{eqnarray}
E&({\bf k})&=0.55J\left( \cos k_x + \cos k_y \right)^2 \nonumber \\
&&+4t'_{eff}\cos k_x\cos k_y +2t''_{eff}\left( \cos 2k_x + \cos 2k_y \right).
\label{Dispersion}
\end{eqnarray}
The first term on the right-hand side is a contribution from $t$ and $J$ terms giving a dispersion from {\bf k}=(0,0) to ($\pi$,$\pi$) with a width of 2.2$J$ and a flat dispersion along the (0,$\pi$)-($\pi$,0) direction.  The second and third terms represent contributions from $t'$ and $t''$, respectively, with effective values $t'_{eff}$ and $t''_{eff}$ proportional to the bare hoppings.  Along the (0,$\pi$)-($\pi$,0) direction, the degeneracy in the first term is lifted by the last two terms.  When the carrier is a hole, the ($\pi$/2,$\pi$/2) point has the lowest energy.  This means that a doped hole goes into ($\pi$/2,$\pi$/2).  On the contrary, a doped electron enters into ($\pi$,0) or (0,$\pi$) when the proper signs of $t'$ and $t''$ are taken into account.  In other words, the band edge of the lower Hubbard band is at ($\pi$/2,$\pi$/2), while that of the upper Hubbard band is at ($\pi$,0).\cite{Tsutsui,Hasan}  It is important to notice that the energy gain due to the $t'$ and $t''$ terms contains information on spin background.\cite{Kim}  Here $t'$ and $t''$ do not change spin configuration because of the same sublattice hoppings.  Therefore the second and third terms in Eq.~(2) are a good measure of the energy of a N\'eel-type spin configuration with a single carrier.  Along the (0,$\pi$)-($\pi$,0) direction, the maximum value of the energy gain due to the terms for a hole is $-$4$|t''_{eff}|$ at {\bf k}=($\pi$/2,$\pi$/2), while that for an electron is $-4(|t'_{eff}|+|t''_{eff}|)$ at {\bf k}=($\pi$,0).  The N\'eel-type spin configuration is thus more stable in the single-electron ground state than in the single-hole ground state, resulting in strong AF correlation for electron doping.

\begin{figure}[t]
\epsfxsize=7.5cm
\centerline{\epsffile{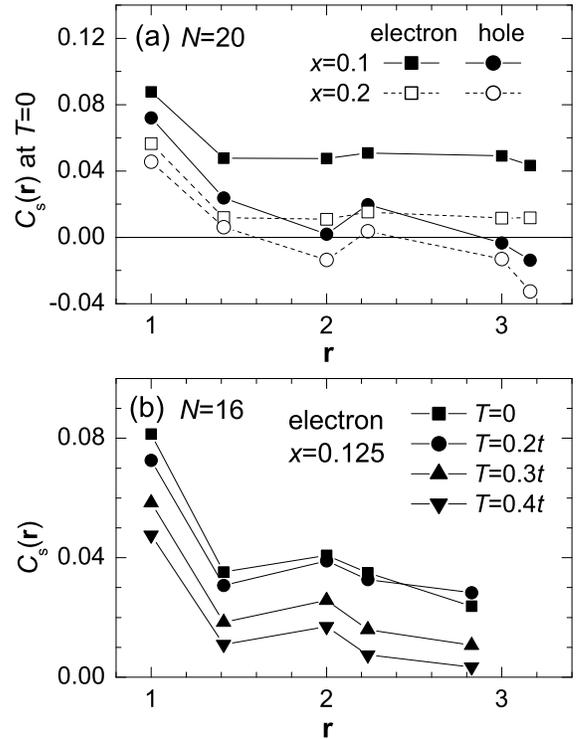}}
\vspace{4mm}
\caption{Staggered spin correlation $C_s({\bf r})$ between spins at site {\bf r} and site 0.  (a) $C_s({\bf r})$ at $T=0$ for a $N$=$\sqrt{20}$$\times$$\sqrt{20}$ $t$-$t'$-$t''$-$J$ cluster with two ($x$=0.1) and four ($x$=0.2) carriers.  $J$/$|t|$=0.4, $t'$/$t$=$-$0.34, and $t''$/$t$=0.23.  Squares and circles denote electron and hole dopings, respectively.  (b) $C_s({\bf r})$ at various temperatures for a $N$=4$\times$4 $t$-$t'$-$J$ cluster with two electrons ($x$=0.125).
}
\label{fig:1}
\end{figure}

The AF correlation remains strong in the underdoped region of the electron-doped $t$-$t'$-$t''$-$J$ model.\cite{Tohyama1,Gooding}  Figure~1(a) shows staggered spin correlations at $T$=0 calculated for a $\sqrt{20}$$\times$$\sqrt{20}$ cluster with periodic boundary conditions (PBC's).  The correlation function is defined as $C_{\rm S}\left({\bf r}\right)=\left<{\rm P}\left({\bf r}\right)S^z_{\bf r} S^z_{\bf 0}\right>$, where $S^z_{\bf r}$ is the $z$ component of the spin operator at site {\bf r}, and ${\rm P}\left({\bf r}\right)$ is 1 when site {\bf r} and site {\bf 0} are in the same sublattice and it is $-$1 otherwise.  In the case of two electrons in the cluster (the concentration $x$=0.1), the AF correlation remains strong at the farthest site, indicating the presence of AF long-range order.  With further doping of electrons (four electrons and $x$=0.2), the AF correlation becomes as weak as those for hole doping.

\begin{figure}[t]
\epsfxsize=7.5cm
\centerline{\epsffile{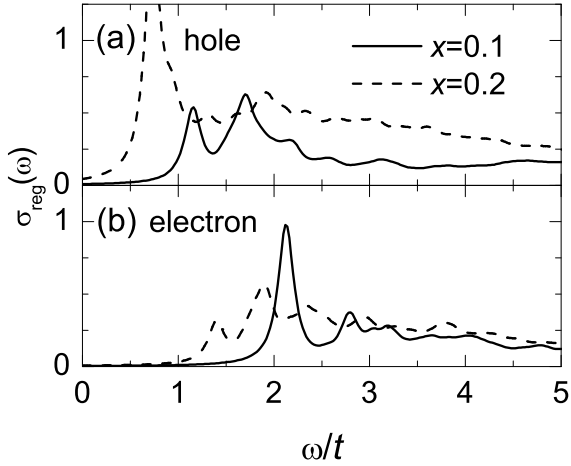}}
\vspace{4mm}
\caption{Regular part of optical conductivity $\sigma_{\rm reg}(\omega)$ at $T$=0 for a $N$=$\sqrt{20}$$\times$$\sqrt{20}$ $t$-$t'$-$t''$-$J$ cluster. (a) Hole doping and (b) electron doping.  Solid and dashed lines represent the carrier concentration of $x$=0.1 and $x$=0.2, respectively.  $J$/$|t|$=0.4, $t'$/$t$=$-$0.34, and $t''$/$t$=0.23.  Delta functions are broadened by a Lorentzian with a width
of 0.1$t$.
}
\label{fig:2}
\end{figure}

We next examine how the AF spin correlation influences charge dynamics.  The real part of optical conductivity [$\sigma_{xx}(\omega)$=$\sigma(\omega)$] is given by the sum of the singular part (the charge stiffness constant or so-called Drude weight $D$) and regular part: $\sigma\left(\omega\right) = 2\pi e^2 D\delta(\omega) + \sigma_{\rm reg} \left(\omega\right)$.  Here $\sigma_{\rm reg}(\omega)$ at $T$=0 for the $\sqrt{20}$$\times$$\sqrt{20}$ $t$-$t'$-$t''$-$J$ cluster is shown in Fig.~2.  The spectrum for electron doping with $x$=0.1 exhibits a large gap structure with $\omega$/$t$$\sim$2 in contract to the hole-doped $t$-$t'$-$t''$-$J$ model.  This gap is very sensitive to not only $J$ but also $t'$ and $t''$: With increasing the absolute values of $t'$/$t$ and $t''$/$t$, the gap increases in energy.  This implies that the gap is closely related to the presence of AF order, since the increase of $t'$/$t$ and $t''$/$t$ stabilizes the AF order.  The gap is thus characterized as an excitation from the AF ground state to an excited state where wrong spin bonds are created by the motion of electrons.  We note that incoherent excitations above the gap are physically the same as those discussed in the context of single-hole motion in the Mott insulator.\cite{Dagotto}  With increasing carrier concentration from $x$=0.1 to $x$=0.2, the spectral weights shift to the lower-energy side and thus the gap value for electron doping decreases, being consistent with the reduction of AF correlation shown in Fig.~1(a).

The AF correlation for electron doping is expected to weaken with increasing temperature.  We thus examine the temperature dependence of spin correlation and charge dynamics.  In the calculations, we employ a 4$\times$4 cluster with two holes or two electrons ($x$=0.125) with PBC's.  Here $t'$/$t$ is set to be $-$0.4, and the $t''$ term is excluded because the cluster is too small to include hoppings connecting sites with two lattice spacing.  In a small system, there is a characteristic temperature below which finite-size effects are appreciable.  This temperature is roughly proportional to an average level spacing in the low-energy sector\cite{Jaklic} and thus $\sim$0.2$t$ in our system. Figure~1(b) shows the temperature dependence of $C_s({\bf r})$ for electron doping.  As expected, the AF correlation rapidly decreases with increasing temperature above $T$=0.2$t$.

\begin{figure}[t]
\epsfxsize=8.0cm
\centerline{\epsffile{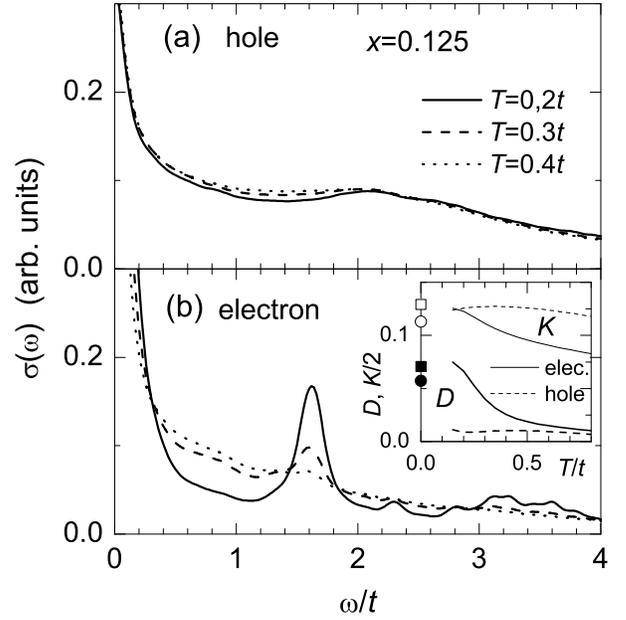}}
\vspace{4mm}
\caption{Temperature dependence of optical conductivity $\sigma(\omega)$ of (a) a hole-doped $t$-$t'$-$J$ model and (b) an electron-doped $t$-$t'$-$J$ model for an $N$=4$\times$4 cluster with two carriers ($x$=0.125).  Here $J$/$|t|$=0.4 and $t'$/$t$=$-$0.4.  A damping factor of 0.1$t$ is introduced both the Drude and regular part of $\sigma(\omega)$.  The inset shows the temperature dependence of $D$ (Drude weight) and $K$ which is roughly proportional to the kinetic energy.  Solid and dashed lines correspond to electron and hole dopings, respectively.  Solid and open squares (circles) denote $D$ and $K$ at $T$=0 for electron (hole) doping.
}
\label{fig:3}
\end{figure}

Figures~3(a) and 3(b) show the temperature dependence of $\sigma(\omega)$, where both the Drude and regular parts are plotted with the same broadening factor of 0.1$t$.  In the hole-doped case, anti-PBC's are adopted along one of the two directions to give a ground state with an absolute energy minimum.  $\sigma(\omega)$ is obtained averaging the two cases (anti-PBC's along the $x$ direction or along the $y$ direction).  $\sigma(\omega)$ for hole doping exhibits a broad distribution of the spectral weight in a wide energy range with small temperature dependence.  In contrast, with decreasing temperature from $T$=0.4$t$ to 0.2$t$, a clear pseudogaplike feature develops for electron doping at $\omega$/$t$$\sim$1 together with a peak structure at $\omega$/$t$=1.6.  The pseudogap is correlated with the development of AF order.  This confirms a picture discussed above that the gap is caused by the presence of the AF order stabilized by long-range hoppings of carriers.  We also note that the temperature at which the pseudogap feature develops is almost one order of magnitude smaller than the gap energy.  This is consistent with experimental data reported recently.\cite{Onose2}

In addition to the gaplike feature, a large Drude contribution is observed near $\omega$=0 in contrast to the hole-doped case.  The Drude weight can be calculated from the sum rule of $\sigma(\omega)$,
\begin{eqnarray}
D=-\frac{K}{2N}-\frac{1}{\pi}\int\nolimits^{\infty}_0\sigma_{\rm reg}\left(\omega\right)d\omega\;,
\label{Drude}
\end{eqnarray}
where $K$=$\langle \tau_{xx}\rangle$, $\tau$ being the stress tensor operator.  In the inset of Fig.~3(b), $D$ and $K$ are plotted as functions of $T$.  Here $D$ is enhanced below $T$/$t$$\sim$0.4 for electron doping.  This is accompanied by an increase of $K$ which reflects an increase of the kinetic energy.  Finite Drude weight at $T$=0 is generally a characteristic signature of a metallic state.  At $T$$>$0, one would expect $D$=0 in the thermodynamic limit of $t$-$J$-type models because there exist scattering processes which prevent the coherent motion of carriers.  Therefore finite $D$ at $T$$>$0 obtained from small systems does not necessarily mean a persist current in macroscopically large systems.\cite{Jaklic}  However, it may be reasonable to regard the finite $D$ observed in our cluster as an indication of a tendency toward a metallic state, because the mean free path for the coherent motion of carriers is, at least, larger than the system size.  It is easy to understand the physical origin of such metallic behavior.  Below $T$/$t$$\sim$0.4=$J$, the AF correlation increases and AF order is established with the help of $t'$ and $t''$.  At the same time, by using $t'$ and $t''$, electron carriers can move in the same sublattice without being disturbed by surrounding spins.  This hoping process induces finite $D$.  In contrast, the spin background for hole doping is strongly disturbed.  Therefore hole carriers are easily scattered by spins, resulting in small $D$ at finite temperature even in our small cluster as shown in Fig.~3(c).

\begin{figure}[t]
\epsfxsize=8.0cm
\centerline{\epsffile{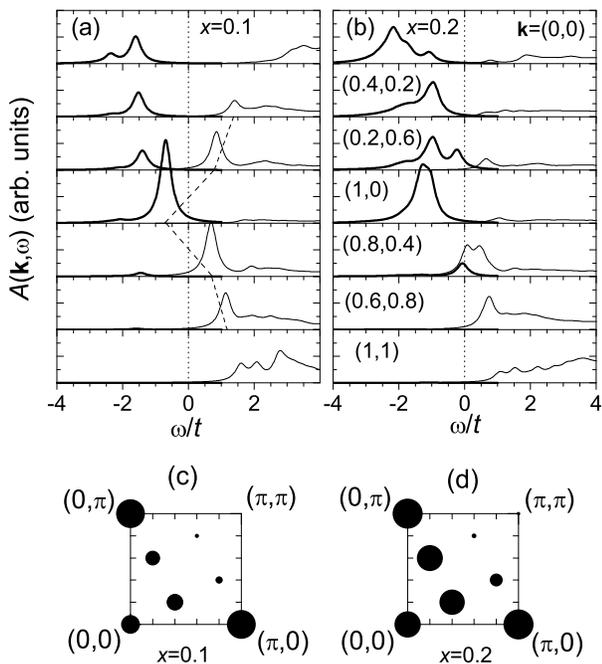}}
\vspace{4mm}
\caption{Single-particle spectral function $A({\bf k},\omega)$ of the electron-doped $t$-$t'$-$t''$-$J$ model for a $\sqrt{20}$$\times$$\sqrt{20}$ cluster.  (a) Two electrons ($x$=0.1) and (b) four electrons ($x$=0.2).  Here $J$/$|t|$=0.4, $t'$/$t$=$-$0.34, and $t''$/$t$=0.29.  Thick and thin curves represent the electron-removal and -addition spectra, respectively, where a Lorentzian broadening with a width of 0.2$t$ is used.  The vertical dotted lines denote the position of the chemical potential.  The momentum is measured in units of $\pi$.  The dashed line in (a) is a guide to the eye.  (c) and (d) show the momentum distribution function $n({\bf k})$ for $x$=0.1 and $x$=0.2, respectively.  The size of solid circles represents the magnitude of $n({\bf k})$.
}
\label{fig:4}
\end{figure}

The spectral function $A({\bf k},\omega)$ also provides useful information on the electronic states.  Figure~4 shows $A({\bf k},\omega)$ at $T$=0 in the electron-doped $\sqrt{20}$$\times$$\sqrt{20}$ $t$-$t'$-$t''$-$J$ cluster.  At $x$=0.1, a large spectral weight exists at {\bf k}=($\pi$,0) in the electron-removal function.  This is a reasonable result because a doped electron at half filling enters into the ($\pi$,0) point as mentioned above.  Since doped electrons are mostly accommodated at ($\pi$,0) and (0,$\pi$) as seen in the momentum distribution function $n({\bf k})$ shown in Fig.~4(c), it is natural to expect an electron pocket showing a small Fermi surface (FS) around {\bf k}=($\pi$,0).  We note that a small amount of spectral weight exists below the Fermi level at (0,0), (0.4$\pi$,0.2$\pi$), and (0.2$\pi$,0.6$\pi$).  At $x$=0.2, doped electrons almost enter into momentum points inside an expected large FS.  Reflecting the large FS, the energy of the ($\pi$,0) quasiparticle becomes deep as compared with that at $x$=0.1.\cite{Kim}  This doping dependence is a great contrast to that in the hole-doped $t$-$J$-type models where a doped hole at half filling enters into ($\pi$/2,$\pi$/2) and at $x$$\sim$0.1 a large FS is already formed.\cite{Stephan}

The paring correlation of doped electrons with $d$-wave symmetry is enhanced when long-range hopping $t'$ with positive sign corresponding to electron doping is introduced into the $t$-$J$ model.\cite{White}  This is accompanied by an enhancement of the AF correlation.  We speculate that the AF correlation exceeds the paring correlation near half filling; i.e., the AF order overcomes the superconducting order.  With increasing electron concentration, the AF correlation weakens and finally the paring correlation may become dominant, resulting in a transition from AF to superconducting order as observed experimentally.  An important point is that the transition may be accompanied by a topology change of the FS from a small surface to a large one as discussed above.

In summary, we have investigated electronic states in the AF phase of the electron-doped cuprates and clarified the nature of the AF phase by employing a numerically exact diagonalization technique for the $t$-$t'$-$t''$-$J$ model.  AF order develops with decreasing temperature with the help of long-range hopping $t'$ and $t''$.  Simultaneously, a gaplike behavior emerges in $\sigma(\omega)$.  There is no such a gap structure in the hole-doped $t$-$t'$-$t''$-$J$ model.  In addition to the gap feature, a large Drude weight is obtained in the small cluster, suggesting coherent motion of the electron carriers.  From examination of the spectral function, we propose that the AF phase in the electron-doped cuprates is characterized as an AF state with small Fermi surface around {\bf k}=($\pi$,0) and (0,$\pi$).  ARPES experiments in the AF phase are desired in order to detect the small Fermi surface.

We would like to thank Y. Onose and Y. Tokura for enlightening discussions.  This work was supported by a Grant-in-Aid for Scientific Research from the Ministry of Education, Culture, Sports, Science and Technology of Japan, and CREST.  One of the authors (S.M.) acknowledges support of the Humboldt Foundation.  The numerical calculations were performed in the supercomputing facilities in ISSP, University of Tokyo, and IMR, Tohoku University.

\medskip

\vfil

\end{document}